\documentclass[pre,aps,twocolumn]{revtex4-2}
\usepackage{amsfonts}
\usepackage{times}
\usepackage{amsmath, amsthm}
\usepackage[colorlinks=true,citecolor=blue]{hyperref}
\usepackage[final]{graphicx}
\usepackage{amssymb, algorithm, algpseudocode}
\algnewcommand\algorithmicforeach{\textbf{for each}}
\algdef{S}[FOR]{ForEach}[1]{\algorithmicforeach\ #1\ \algorithmicdo}
\usepackage[title]{appendix}
\usepackage{xcolor,comment}

\newcommand{\sech}{\mathrm{sech}}

\DeclareMathOperator*{\argmax}{arg\,max}

\begin{document}

\title{Localized patterns and dispersive structures in two-dimensional Fermi-Pasta-Ulam lattices}

\author{Su Yang}
\email{Corresponding author: suyangchina321611@gmail.com}

\affiliation{College of Mathematics and Statistics, Chongqing University, Chongqing 401331, China}

\author{Wenrong Sun}
\affiliation{School of Mathematics and Physics, University of Science
and Technology Beijing, Beijing 100083, China}

\date{\small\today}

\begin{abstract}
    In this paper, we study an analog of the scalar two-dimensional Fermi-Pasta-Ulam (FPU) lattice. In particular, a variety of dispersive wave structures and localized patterns are numerically identified in the numerical simulations of the FPU lattice, but, to the best of our knowledge, all of these particular wave structures do not admit analytical closed-form expressions. In order to resolve this issue, we perform a dimensional reduction and accordingly derive a modified KdV equation. Based on this reduction, we first take advantage of some of its exact localized solutions to model the associated wave patterns in the FPU lattice. In addition, we explore the two-dimensional generalization of the Riemann problems for the FPU lattice and the corresponding modified KdV reduction, whose evolution dynamics lead to the formation of multiple composite dispersive structures. Moreover, we propose and rigorously derive the KPII limit of the FPU lattice and investigate their associated wedge problems. Finally, all these relevant numerical dynamics are compared to examine the performance of these quasi-continuum long-wave asymptotic limits.  
\end{abstract}

\maketitle

\section{Introduction}

Rogue waves (RWs) \cite{kharif2008rogue,pelinovsky2008extreme} are classified as extreme wave phenomena that show up nowhere and disappear with no trace. RWs also admit prototypical characteristics that their amplitudes are at least twice greater than the significant wave height \cite{charalampidis2018phononic}. The mathematical foundations of RWs can be traced back to the seminal work of Peregrine \cite{peregrine1983water} where the so-called Peregrine soliton solution to the nonlinear Schr\"odinger equation has been rigorously derived to model the spatial profiles of RWs. In addition, recent studies on the modified Korteweg–De Vries (mKdV) equation have also revealed the existence of RWs \cite{chen2018rogue,ankiewicz2018rogue,ankiewicz2019shallow}. Interestingly, these extreme wave events are experimentally observed in the setting of Bose-Einstein condensates \cite{romero2024experimental,bludov2009matter,bludov2010vector,manikandan2014manipulating}. On the other hand, nonlinear dispersive wave patterns are typical coherent structures in a variety of models of mathematical physics. Dispersive shock waves (DSWs) \cite{hoefer2009dispersive} are non-stationary and smooth multi-scale pattern that connects states with distinct amplitudes by an expanding wavetrain. Their investigations have been conducted since the past five decades, which are initiated by the pioneering works of Gurevich \& Pitaevskii \cite{gurevich1973nonstationary} and Whitham \cite{whitham2011linear}, and their theoretical foundations are recently summarized in review articles \cite{el2016dispersive,el2017dispersive}. DSWs and analogous wave patterns can be numerically observed in multiple dispersive models such as the nonlinear Schr\"odinger equation \cite{mohapatra2026dam,yang2026dispersive,chandramouli2024dispersive} and granular crystal lattices \cite{yang2025regularized,yang2026first,yang2026quasicontinuum}. Moreover, their experimental existence can be leveraged in settings such as plasmas and superfluids \cite{maiden2016observation,xu2017dispersive}. The mathematical analysis of DSWs are on the basis of the Whitham modulation theory \cite{abeya2023whitham,ablowitz2017whithamBBM,biondini2024whitham,whitham2011linear,ablowitz2017whithamKP} which shall yield a system of partial differential equations (PDEs) that govern the slowly varying spatial and temporal dynamics of the parameters of the periodic traveling waves. Importantly, the Whitham modulation system is shown to admit two useful reductions at its linear and soliton limits \cite{el2005resolution,el2016dispersive}. The two reduced modulation system forms the so-called ``simple-wave" ordinary differential equations (ODEs) whose solutions leverage quantitative predictions on multiple edge features of DSWs, and the reductions of the Whitham system are named ``DSW fitting" \cite{el2016dispersive}.

In the present work, we are interested in studying both extreme and dispersive wave structures in an analog of the two-dimensional FPU lattice. The FPU lattice has a wide physical applications in particle physics. For instance, the so-called granular crystals or granular chain is one particular type of the FPUT system, which models the interaction dynamics between closely packed particles. Previous studies in Refs.~\cite{charalampidis2018phononic,miyazawa2022rogue,yasuda2017emergence} have explored and shown the existence of RWs and DSWs in the spatially one-dimensional FPU lattice. To the best of our knowledge, none of these previous works has considered the spatially two-dimensional FPU lattice and examined the existence of extreme and dispersive structures therein. Hence, the lack of investigations on these pertinent wave phenomena in two-dimensional FPU lattices becomes the natural motivation of the present work. In particular, we shall first perform a dimension reduction and derive an associated modified KdV reduction of the analog of the two-dimensional FPUT lattice. This particular quasi---continuum reduction will be utilized to explore ``line" RWs phenomena in the lattice, which is similar to the line soliton of the Kadomtsev–Petviashvili (KP) equation \cite{biondini2026kinetic,PhysRevResearch.7.013143,biondini2007line}. Meanwhile, we also apply this mKdV reduction to approximate and describe the regular ``line" DSW and other composite dispersive patterns specified in Ref.~\cite{el2017dispersive}. Furthermore, we compute the two-dimensional KP asymptotic limit of the lattice. With the KP reduction, we probe the wedge problems dynamics \cite{cdvf-xnfw} corresponding to the original lattice system with consistent initial conditions.

This work is structured as follows. In Sec.~\ref{Sec: Intro}, we give an overview of the lattice of interest for this paper and elaborate on some of its particular features. We then rigorously derive these two asymptotic reductions of mKdV and KPII equations that will be used to study specific wave phenomena in the main lattice model. In Sec.~\ref{Sec: Initial conditions set-up}, we discuss some necessary numerical preliminaries including the numerical schemes utilized to simulate all the relevant models and the detailed procedures to construct the initial data for each models. Next, in Sec.~\ref{Sec: RWs explore}, we first list some particular exact solutions of the mKdV reduction which can be used to approximate the ``line" rogue waves of the original lattice. In the same section, we also perform numerical studies on the ``line" dispersive patterns of the lattice with the approximations of the mKdV reduction. In Sec.~\ref{Sec: Wedge problems}, we apply the KPII asymptotic limit to investigate the wedge problems of the lattice model. Finally, the paper ends with an overall summary of this work and some open questions for future research directions in Sec.~\ref{Sec: conclusions}.

\section{Model description and theoretical setup}\label{Sec: Intro}

In the present work, we are interested in the following (1+2) second-order (in time) two-dimensional lattice \cite{chong2026travelingdispersiveshockwaves}:
\begin{equation}\label{Main model}
    \frac{d^2u_{n,m}}{dt^2} = \Delta_d\Phi(u),
\end{equation}
where $u_{n,m}(t):\mathbb{Z}^2\times\mathbb{R}\to\mathbb{R}$ is the field variable, $\Phi$ is assumed to be a convex function, and $\Delta_d$ refers to the discrete Laplacian operator which is defined as follows: For any function $f$,
\begin{equation}\label{Laplace operator}
    \Delta_d(f) \equiv f_{n+1,m}+f_{n-1,m}+f_{n,m+1}+f_{n,m-1}-4f_{n,m}.
\end{equation}
We notice first that Eq.~\eqref{Main model} represents the central-difference discretization of the following dispersionless continuum system:
\begin{equation}
    u_{tt} = \Delta\left(\Phi(u)\right), \quad \Delta = \partial^2_x + \partial^2_y.
\end{equation}
Throughout this paper, for illustrative purposes, we assume that $\Phi(r) = K_1r + K_2r^2 + K_3r^3$, where $K_{1,2,3} \in \mathbb{R}$ are three real constants.

Then, we look for a plane-wave solution in the form of an infinitesimal perturbation near the homogeneous state of $\overline{u}$: $u_{n,m}(t) = \overline{u} + \epsilon\exp\left[i\left(kn+lm-\omega t\right)\right]$, where $0<\epsilon\ll 1$. Substituting such plane-wave ansatz into Eq.~\eqref{Main model} yields the following linear dispersion relation, upon the elimination of the smallness parameter of $\epsilon$:
\begin{equation}\label{ldr of the Main model}
    \omega^2(k,l;\overline{u}) = 4\left(K_1+2\overline{u}K_2\right)\left(\sin^2\left(\frac{k}{2}\right) + \sin^2\left(\frac{l}{2}\right)\right).
\end{equation}
Then, it is worthwhile to notice that $\partial^2_{k,l}\omega_+ \neq 0$ for $k,l \neq 2m\pi$, $m\in\mathbb{Z}$. Hence, the lattice in Eq.~\eqref{Main model} is dispersive, and we shall expect the emergence of dispersive wave patterns of Eq.~\eqref{Main model} via performing time stepping for appropriate initial conditions.

\section{Long-wave asymptotic limits}\label{Sec: Quasicontinuum reductions}

In this section, we derive the two relevant long-wave quasi-continuum models, the modified KdV and KPII equations, which shall be applied to approximate numerous wave structures in the two-dimensional FPUT lattice in Eq.~\eqref{Main model}. However, before proceeding to the asymptotic reductions, it is important to clarify the physical role of the quadratic coefficient $K_2$ in the onsite potential $\Phi(u)=K_1u+K_2u^2+K_3u^3$.
The equation of motion \eqref{Main model} is invariant under the inversion transformation $u_{n,m}\to -u_{n,m}$ if and only if $\Phi(u)$ is an odd function, which requires all even-order coefficients to vanish.
Hence, the choice $K_2=0$ corresponds to a lattice that possesses full inversion symmetry, i.e., the restoring force is symmetric with respect to the sign of the displacement.
In this case, the lowest-order nonlinearity is cubic, and the long-wave asymptotics is governed by the modified KdV equation derived in Sec.~\ref{Subsec: mKdV}.
Conversely, when $K_2\neq 0$, the inversion symmetry is broken: the potential is no longer odd, and the lattice responds differently to positive and negative displacements.
The quadratic nonlinearity then dominates the weakly nonlinear dynamics, and the appropriate asymptotic model is the Kadomtsev--Petviashvili equation, which retains the quadratic term (Sec.~\ref{Subsec: KPII reduction}).
In what follows, we treat the two scenarios separately: the mKdV reduction for inversion-symmetric lattices ($K_2=0$), and the KP reduction for asymmetric lattices ($K_2\neq 0$).

\subsection{Modified KdV reduction}\label{Subsec: mKdV}

Based on Ref.~\cite{chong2026travelingdispersiveshockwaves}, with the assumption that $K_2 = 0$, if we consider the asymptotic slow variables:
\begin{equation}\label{mKdV cov}
   \begin{aligned}
    &u_{n,m}(t) = \epsilon U(X,T),\\
    &X = \epsilon\left(n+m-c_st\right), \quad T = \epsilon^3 t,
    \end{aligned}
\end{equation}
where $c_s = \sqrt{2K_1}$ refers to the speed of sound. 

Substitution of Eq.~\eqref{mKdV cov} into Eq.~\eqref{Main model} yields the following modified KdV equation on the order of $\mathcal{O}(\epsilon^5)$, upon the integration with respect to the spatial variable of $X$:
\begin{equation}\label{mKdV reduction}
    U_T + \frac{3K_3}{c_s}U^2U_X + \frac{K_1}{12c_s}U_{XXX} = 0.
\end{equation}

\subsection{KPII reduction}\label{Subsec: KPII reduction}

Now, we rigorously derive the Kadomtsev-Petviashvili (KP) asymptotic reduction of Eq.~\eqref{Main model}. We utilize the following multi-scale change of variables:
\begin{equation}\label{multi-scale c.o.v}
   \begin{aligned}
    &u_{n,m}(t) = \epsilon^2U(X,Y,T),\\
    &X = \epsilon(n - \tilde{c}_st),\quad Y=\epsilon^2m, \quad T = \epsilon^3t,
    \end{aligned}
\end{equation}
where $\tilde{c}_s^2 = K_1$ denotes the sound speed.

Substituting Eq.~\eqref{multi-scale c.o.v} into Eq.~\eqref{Main model} and collecting relevant terms on the order of $\mathcal{O}(\epsilon^6)$ yields the following KPII equation:
\begin{equation}\label{KP reduc}
    \left(U_T+\frac{K_2}{\tilde{c}_s}UU_X+\frac{K_1}{24\tilde{c}_s}U_{XXX}\right)_X + \frac{K_1}{2\tilde{c}_s}U_{YY} = 0.
\end{equation}
Furthermore, through the scalings of $\tilde{X} = \frac{2\sqrt{6}\sqrt{K_2}}{\sqrt{K_1}}X$, $\tilde{Y} = \frac{96^{1/4}K_2^{3/4}}{K_1^{3/4}}Y$, and $\tilde{T} = \frac{K_2}{\tilde{c}_s}T$, we can cast all the coefficients of KPII reduction in Eq.~\eqref{KP reduc} to unity:
\begin{equation}\label{KPII with unity}
    \left(U_{\tilde{T}} + UU_{\tilde{X}} + U_{\tilde{X}\tilde{X}\tilde{X}}\right)_{\tilde{X}} + U_{\tilde{Y}\tilde{Y}} = 0.
\end{equation}

\section{Numerical scheme \& initial conditions}\label{Sec: Initial conditions set-up}

Before we dive into the exploration of the wave patterns of all the relevant models, we will discuss in detail the numerical methods utilized to solve these three models in Eqs.~\eqref{Main model}, \eqref{mKdV reduction}, and \eqref{KPII with unity}.

For the two quasi-continuum models in Eqs.~\eqref{mKdV reduction} and \eqref{KPII with unity}, we apply the ETDRK4 time integration scheme \cite{kassam2005fourth} with pseudo-spectral discretization of the space to numerically solve them. Since they are both first-order uni-directional model, we only need an initial data for the field variable of $U$, and denote such initial conditions at $t = t_0$ as $U^0$ and $\tilde{U}^0$ for Eqs.~\eqref{mKdV reduction} and \eqref{KPII with unity}, respectively.

Now, to numerically solve the FPUT lattice in Eq.~\eqref{Main model}, we cast it into the following equivalent first-order system:
\begin{equation}\label{Equivalent first-order system}
    \begin{aligned}
        &\frac{du_{n,m}}{dt} = v_{n,m},\\
        &\frac{dv_{n,m}}{dt} = \Delta_d\Phi(u),
    \end{aligned}
\end{equation}
where $v_{n,m}(t)$ is defined as the velocity variable. On the one hand, we note that the initial data for $u_{n,m}$ at $t = t_0$, which is consistent with $U^0$ and $\tilde{U}^0$ utilized for Eqs.~\eqref{mKdV reduction} and \eqref{KPII with unity} simply read:
\begin{equation}\label{ICs for the lattice}
   \begin{aligned}
    &u_{n,m}(t_0) = \epsilon U^0\left(\epsilon (n+m-c_st_0), \epsilon^3t_0\right),\\
    &u_{n,m}(t_0) = \epsilon^2\tilde{U}^0\left(\epsilon\left(n-c_st_0\right),\epsilon^2m,\epsilon^3t_0\right),
    \end{aligned}
\end{equation}
respectively.

On the other hand, for the velocity variable of $v_{n,m}(t)$, we observe first that, by the chain rules, associated with the mKdV equation \eqref{mKdV reduction}:
\begin{equation}
    \begin{aligned}
        v_{n,m}(t) &= \frac{du_{n,m}}{dt} \\
    &=\epsilon\partial_tU\\
    &= -c_s\epsilon^2U_X + \epsilon^4U_T.
    \end{aligned}
\end{equation}
Then, corresponding to the KPII equation \eqref{KPII with unity}, we have that
\begin{equation}\label{ICs for velocity associated with KPII}
   \begin{aligned}
    &v_{n,m}(t) = -a\tilde{c}_s\epsilon^3U_{\tilde{X}} \\
    &- c\epsilon^5\left(UU_{\tilde{X}}+U_{\tilde{X}\tilde{X}\tilde{X}}+\partial_{\tilde{Y}}^2\int^{\tilde{X}}U(X')dX'\right),
    \end{aligned}
\end{equation}
where we have suppressed the independent variables of $\tilde{Y}$ and $\tilde{T}$ for notational simplicity, and note that $a = \frac{2\sqrt{6}\sqrt{K_2}}{\sqrt{K_1}}$ and $c = \frac{K_2}{\tilde{c}_s}$.

Hence, the initial data which are consistent to Eqs.~\eqref{mKdV reduction} and \eqref{KPII with unity} for $v_{n,m}$ at $t = t_0$ read:
\begin{equation}\label{IC for velocity}
   \begin{aligned}
    &v_{n,m}(t_0) = -c_s\epsilon^2U^0_X + \epsilon^4U^0_T,\\
    &v_{n,m}(t_0) = -a\tilde{c}_s\epsilon^3\tilde{U}^0_{\tilde{X}} \\
    &- c\epsilon^5\left(\tilde{U}^0\tilde{U}^0_{\tilde{X}} + \tilde{U}^0_{\tilde{X}\tilde{X}\tilde{X}} + \partial^2_{\tilde{Y}}\int^{\tilde{X}}\tilde{U}^0(X')dX'\right),
    \end{aligned}
\end{equation}
respectively.

\section{Localized patterns via mKdV asymptotics}\label{Sec: RWs explore}

In this section, we shall use some existing and known localized patterns from the mKdV reduction \eqref{mKdV reduction} to model the rogue wave phenomena in the lattice equation \eqref{Main model}. Specifically, we apply several analytically exact solutions of Eq.~\eqref{mKdV reduction} to set up proper initial conditions for Eq.~\eqref{Main model} and examine their evolution dynamics.

\subsection{Rational solutions and rogue waves}
The mKdV reduction in Eq.~\eqref{mKdV reduction} admits multiple exact rational solutions which can be utilized to model the profiles of rogue waves. In particular, on the one hand, it possesses the following second-order solution \cite{ankiewicz2018rogue} denoted as $U_2$:
\begin{equation}\label{mKdV 2nd-order soln}
    U_2 = \frac{12G_2}{D_2} + 1,
\end{equation}
with 
\begin{equation}
   \begin{aligned}
    &G_2 = 3 - \left(6aT + bX\right)\left[\left(6aT+bX\right)^3 + 6\left(22aT + bX\right)\right],\\
    &D_2 = 12aT\bigg[243(2a)^4T^4\left(aT+bX\right)\\
    &+bX\left(3b^4X^4-2b^2X^2+51\right)\\
    &+72a^2bXT^2\left(5b^2X^2-9\right)\\
    &+108a^3T^3\left(15b^2X^2-13\right)\\
    &+3aT\left(15b^4X^4-30b^2X^2+139\right)\bigg] \\
    &+b^6X^6+3b^4X^4+27b^2X^2+9,
    \end{aligned}
\end{equation}
where
\begin{equation}\label{coefficients for 2nd order soln}
        a = \frac{b^3\gamma_3}{4}, \quad b = \sqrt{\frac{24K_3}{K_1}}, \quad
        \gamma_3 = -\frac{K_1}{12c_s}.
\end{equation}

\begin{figure*}
    \centering
    \includegraphics[width=0.8\linewidth]{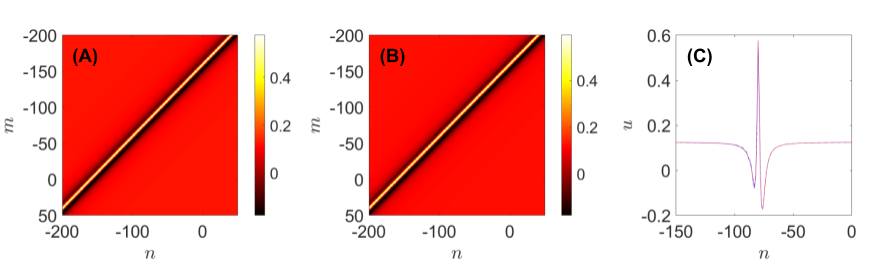}
    \hfill
    \includegraphics[width=0.8\linewidth]{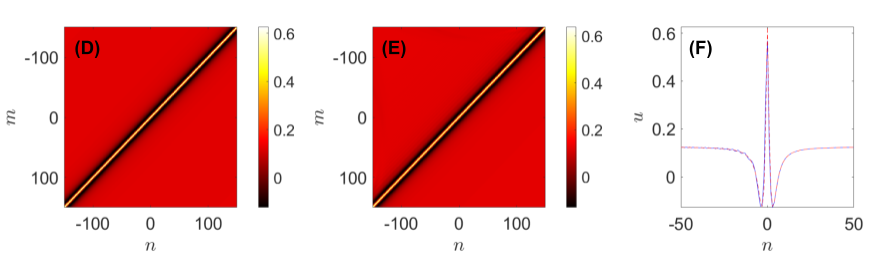}
    \hfill
    \includegraphics[width=0.8\linewidth]{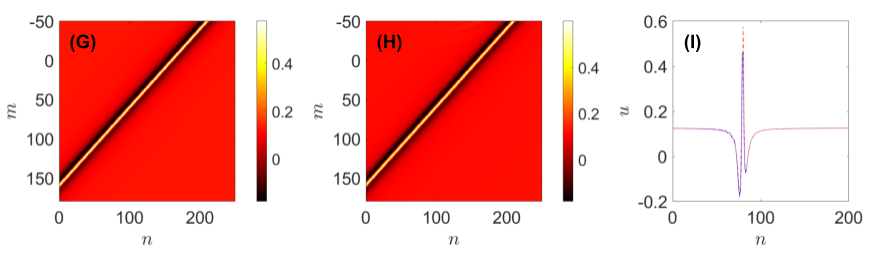}
    \caption{Comparison of the second-order rational solution in Eq.~\eqref{mKdV 2nd-order soln}. The leftmost, middle, and rightmost columns represent the analytical, numerical, and cross-sections along the direction of $m = n$ solutions. These comparisons are performed at $t = -50$ (Panels (A)-(C)), $t = 0$ (Panels (D)-(F)), and $t = 50$ (Panels (G)-(I)), respectively. Notice that $\epsilon = 0.125$, $K_1 = 5$, and $K_3 = 1$ in all the relevant numerical experiments.}
    \label{fig:2nd-order solution comparison}
\end{figure*}

Fig.~\ref{fig:2nd-order solution comparison} depicts the numerical comparisons between the second-order rational solution in Eq.~\eqref{mKdV 2nd-order soln} with the associated numerical dynamics of the lattice \eqref{Main model}. In particular, the respective three row shows the dynamics of Eq.~\eqref{Main model} at $t = -50, 0, 50$. Based on the cross-sectional profile comparisons in Panels (C), (F), and (I), we can see that the numerical dynamics agrees very well with the analytical counterparts.

On the other hand, Eq.~\eqref{mKdV reduction} also admits the following third-order solution \cite{ankiewicz2018rogue} denoted as $U_3$:
\begin{equation}\label{3rd-order rational soln}
    U_3 = \frac{24G_3}{D_3} - 1,
\end{equation}
with
\begin{equation}
    \begin{aligned}
        &G_3 = 4\bigg[800c^3t^3z + 150c^2t^2\left(16z^4-8z^2+5\right) \\
        &+120ctz\left(-16z^4+40z^2+15\right)\\
        &+z^2\left[8z^2\left(32z^6+120z^4+420z^2-225\right)-675\right]\bigg] + 675,\\
        &D_3 = 2025 + 8\bigg[800c^4t^4-800c^3t^3z\left(-3+4z^2\right)\\
        &+30c^2t^2\left(165-180z^2+240z^4+64z^6\right)\\
        &+10ctz\left[-675+32z^2\left(45-27z^2+8z^6\right)\right] + z^2\times\\
        &\left[6075+2z^2\left(3375+16z^2\left[585+135z^2+24z^4+16z^6\right]\right)\right]\bigg],
    \end{aligned}
\end{equation}
where
\begin{equation}
    z = \frac{1}{4}\left(2bX - cT\right), \quad c = \frac{b^3K_1}{4c_s}.
\end{equation}

\begin{figure*}
    \centering
    \includegraphics[width=0.8\linewidth]{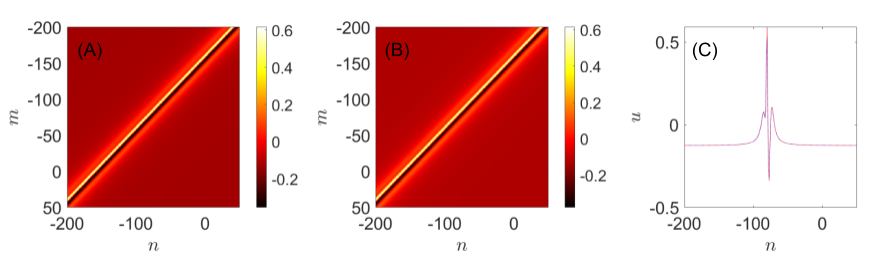}
    \hfill
    \includegraphics[width=0.8\linewidth]{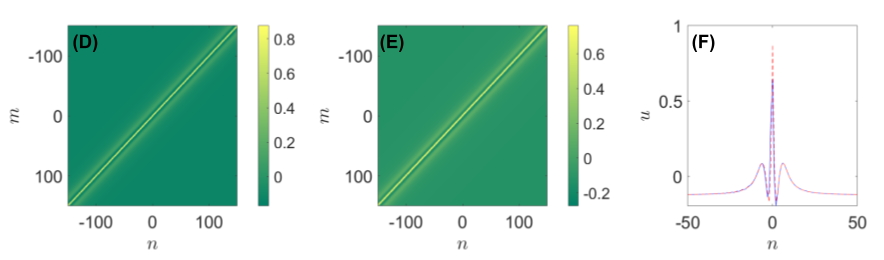}
    \hfill
    \includegraphics[width=0.8\linewidth]{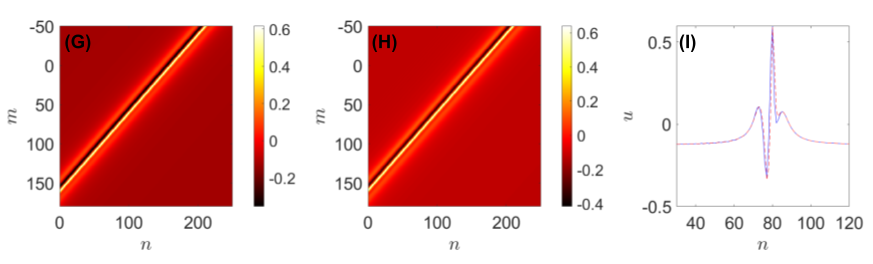}
    \caption{Same as Fig.~\ref{fig:2nd-order solution comparison} but now for the third-order rational solutions specified in Eq.~\eqref{3rd-order rational soln}.}
    \label{fig:3rd-order rational solution comparison}
\end{figure*}

Fig.~\ref{fig:3rd-order rational solution comparison} displays the numerical comparisons of the ``line" rogue waves modeled by the third-order rational solution specified in Eq.~\eqref{3rd-order rational soln}. These comparisons are analogous to those of the second-order rational solutions, which demonstrate to be equally good.

\section{Dispersive structures}

\begin{figure}[b!]
    \centering
    \includegraphics[width=0.99\linewidth]{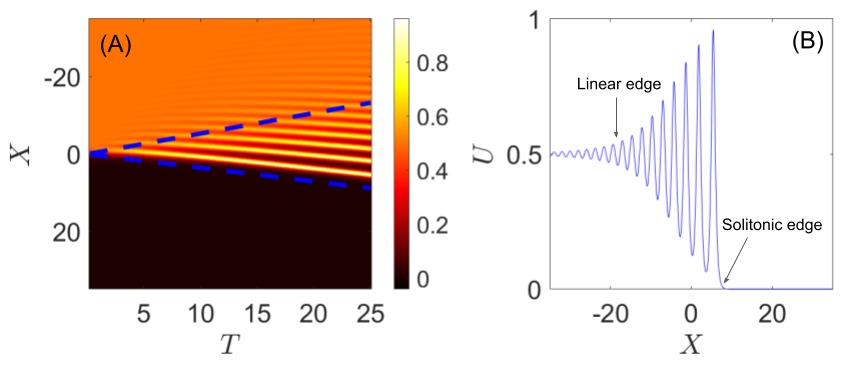}
    \caption{The spatiotemporal dynamics (A) and the spatial profile (B) of the regular DSW of the mKdV equation \eqref{mKdV reduction}. Notice that the two blue dashed lines in Panel (A) represent the DSW fitting predictions on the linear and solitonic locations of the mKdV DSW, which are essentially the two lines of $X = S_{\pm}^{(X,T)} T$, where $S_{\pm}^{(X,T)}$ are given in Eq.~\eqref{edge speeds in (X,T)}.}
    \label{fig:regular DSW}
\end{figure}

Next, we investigate dispersive structures from the pertinent numerical simulations of Eq.~\eqref{Main model}. These dispersive phenomena will include the regular DSW and other compound waves. We also apply the typical DSW fitting method to the mKdV equation \eqref{mKdV reduction} to obtain different theoretical quantitative insights on the edge characteristics of the regular DSW.

\subsection{Dispersive shocks}

\begin{figure}[t!]
    \centering
    \includegraphics[width=0.99\linewidth]{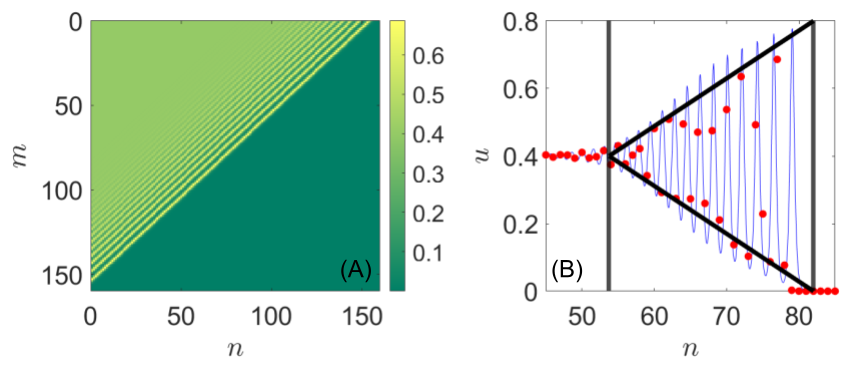}
    \caption{The comparison of the regular DSW wave profiles between the FPUT lattice \eqref{Main model} and its modified KdV reduction \eqref{mKdV reduction} at $t = 100$. Panel (A) displays the density plot of the numerical DSW of the lattice \eqref{Main model}, while (B) depicts the cross-sections of the numerical solutions of Eq.~\eqref{Main model} (red dots) and Eq.~\eqref{mKdV reduction} (blue curve) along the direction of $m = n$.}
    \label{fig: regular DSW comparison}
\end{figure}

In this section, we shall utilize the class of novel dispersive patterns of the mKdV reduction in Eq.~\eqref{mKdV reduction} to explore the corresponding structures in the FPUT lattice \eqref{Main model}. The prototypical medium to investigate dispersive structures is the so-called Riemann problem associated with the model of interests. In particular, we consider the mKdV equation in Eq.~\eqref{mKdV reduction} with the following steplike initial condition:
\begin{equation}\label{Riemann Initial data}
    U(X,0) = \begin{cases}
        U_-, \quad X \leq 0,\\
        U_+, \quad X>0,
    \end{cases}
\end{equation}
where $U_{\pm} \in \mathbb{R}$ are two parameters which represent the homogeneous background.
Fig.~\ref{fig: regular DSW comparison} depicts the profiles of the mKdV DSW evolved from the Riemann initial data \eqref{Riemann Initial data} with $U_- = 0.5$ and $U_+ = 0$. Notice that the respective linear and solitonic edge of the mKdV DSW is labeled in Panel (B).

Due to the non-convexity feature of the nonlinearity of the mKdV reduction \eqref{mKdV reduction}, the evolution dynamics of the Riemann initial data can result in a variety of interesting dispersive wave phenomenon including the regular DSWs, and composite dispersive waves such as kink-DSW and rarefaction-DSW. 

\begin{figure}[b!]
    \centering
    \includegraphics[width=0.99\linewidth]{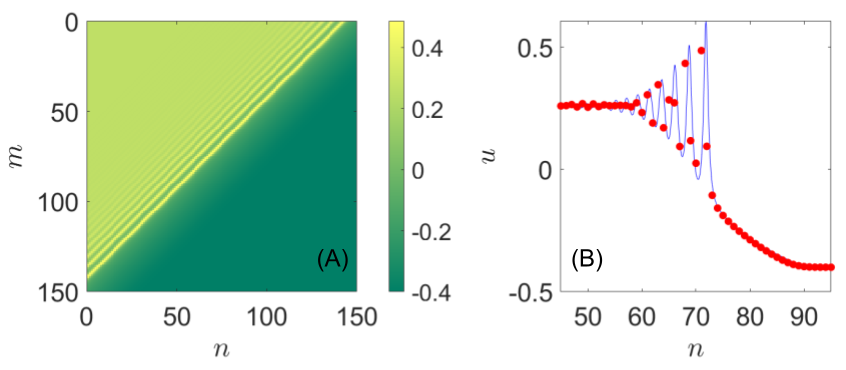}
    \caption{Same as Fig.~\ref{fig: regular DSW comparison} but now for the rarefaction-DSW.}
    \label{fig:R-DSW comparison}
\end{figure}

\begin{figure}[t!]
    \centering
    \includegraphics[width=0.99\linewidth]{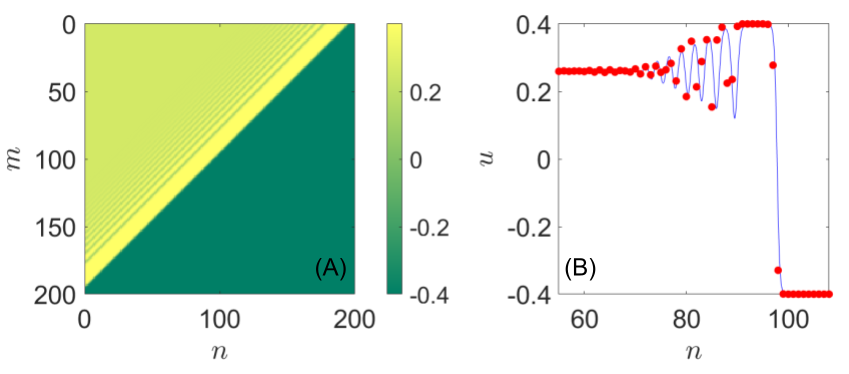}
    \caption{Same as Fig.~\ref{fig: regular DSW comparison} but now for the Kink-DSW at $t = 150$.}
    \label{fig:Kink-DSW comparison}
\end{figure}

\subsection{DSW fitting \& edge-feature predictions}

Interestingly, the edge speeds of the DSW of the modified KdV equation \eqref{mKdV reduction} including its linear and solitonic-edge speeds can be theoretically predicted by the method of DSW fitting. This particular method stems from the Whitham modulatin system associated with the dispersive models of interest, which leverages the slowly varying evolution dynamics of all the relevant parameters of the periodic traveling waves. Furthermore, at both the harmonic and solitonic limits of the modulation equations, the Whitham system admits two very useful reductions. In particular, these reductions are represented by two initial-value problems (IVPs) which encode important theoretical insights on the edge characteristics of the DSWs. These two IVPs read \cite{el2016dispersive}:
\begin{equation}\label{Two simple-wave ODEs}
    \begin{aligned}
        &\frac{dK}{d\overline{U}} = \frac{\partial_{\overline{U}}\Omega_0}{V(\overline{U}) - \partial_{K}\Omega_0}, \quad K(U_+) = 0,\\
        &\frac{d\widetilde{K}}{d\overline{U}} = \frac{\partial_{\overline{U}}\Omega_s}{V(\overline{U}) - \partial_{\widetilde{K}}\Omega_s},\quad \widetilde{K}(U_-) = 0.
    \end{aligned}
\end{equation}
We notice that $K,\widetilde{K},\overline{U}$ in Eqs.~\eqref{Two simple-wave ODEs} are defined as the wavenumber, conjugate wavenumber, and the averaged homogeneous background, and $\Omega_0,\Omega_s$ refer to the linear and conjugate dispersion relations. In particular, for the mKdV reduction in Eq.~\eqref{mKdV reduction}, its linear dispersion relation of $\Omega_0$ can be derived by firstly looking for a plane-wave ansatz in the form of $U(X,T) = \overline{U} + \eta\exp\left[i(KX-\Omega T)\right]$, where $ 0 < \eta \ll 1$. Substitution of this ansatz into Eq.~\eqref{mKdV reduction} yields, by eliminating the smallness parameter of $\eta$:
\begin{equation}
    \Omega_0(K,\overline{U}) = \frac{3K_3}{c_s}\overline{U}^2K - \frac{K_1}{12c_s}K^3,
\end{equation}
and the conjugate dispersion relation can be accordingly obtained as follows:
\begin{equation}
    \Omega_s(\widetilde{K},\overline{U}) = -i\Omega_0(i\widetilde{K},\overline{U}).
\end{equation}
Moreover, the dispersionless velocity of the mKdV reduction \eqref{mKdV reduction} reads $V(\overline{U}) = \frac{3K_3}{c_s}\overline{U}^2$.

Solving these two IVPs in Eqs.~\eqref{Two simple-wave ODEs} yields the following solutions for $K$ and $\widetilde{K}$:
\begin{equation}\label{solns for wavenumbers}
    \begin{aligned}
        &K_-^2 = \frac{24K_3}{K_1}\left(U_-^2 - U_+^2\right),\\
        &\widetilde{K}_+^2 = \frac{24K_3}{K_1}\left(-U_+^2+U_-^2\right).
    \end{aligned}
\end{equation}
Then, the linear and solitonic-edge speeds, denoted as $S_-$ and $S_+$, of the DSW of the mKdV reduction are obtained by computing the following group and phase velocity, respectively:
\begin{equation}\label{edge speeds in (X,T)}
    \begin{aligned}
        &S_-^{(X,T)} = \partial_K\Omega_0\left(K_-,U_-\right) = \frac{3K_3}{c_s}U_-^2 - \frac{6K_3}{c_s}\left(U_-^2 - U_+^2\right),\\
        &S_+^{(X,T)} = \frac{\Omega_s}{\widetilde{K}}\left(\widetilde{K}_+, U_+\right) = \frac{3K_3}{c_s}U_+^2 + \frac{2K_3}{c_s}\left(-U_+^2 + U_-^2\right).
    \end{aligned}
\end{equation}
It is worth noting that the DSW-fitting theoretical predictions on the edge speeds in Eq.~\eqref{edge speeds in (X,T)} need to fulfill the following ``entropy conditions" \cite{el2005resolution,el2016dispersive}:
\begin{equation}\label{Entropy conditions}
    \begin{aligned}
        S_-^{(X,T)} < V(U_-), \quad S_+^{(X,T)} > V(U_+), \quad S_-^{(X,T)} < S_+^{(X,T)},
    \end{aligned}
\end{equation}
where $V(U_{\pm}) = \frac{3K_3}{c_s}U_{\pm}^2$. The entropy conditions \eqref{Entropy conditions} shall specify a parametric regime for the two parameters of $U_{\pm}$, which ensures the applicability of DSW fitting. For instance, for the values of $U_{\pm}$ lead to the formation of the Kink-DSW delineated in Fig.~\ref{fig:Kink-DSW comparison}, the entropy conditions specified in Eq.~\eqref{Entropy conditions} will be clearly violated and the theoretical DSW-fitting predictions on edge features of the DSW become invalid.

Furthermore, we note that the super-script of $(X,T)$ utilized in Eqs.~\eqref{edge speeds in (X,T)} emphasizes the fact that these two edge speeds represent the theoretical predictions in the $(X,T)$ coordinates. Since our main model \eqref{Main model} of interests is considered in the coordinates of $(n,m,t)$, it is necessary for transform these predictions \eqref{edge speeds in (X,T)} to their counterparts within $(n,m,t)$. To this end, we apply the relations specified in Eq.~\eqref{mKdV cov} and observe that
\begin{equation}\label{bridge bettwen (n,m,t) and (X,T)}
    \frac{d(n+m)}{dt} = c_s + \epsilon^2\frac{dX}{dT}.
\end{equation}
Throughout this work, since we are interested only in considering comparisons along the direction of $m = n$ so that Eq.~\eqref{bridge bettwen (n,m,t) and (X,T)} can be further simplified as follows:
\begin{equation}\label{key relation}
    \frac{dn}{dt} = \frac{1}{2}\left(c_s + \epsilon^2\frac{dX}{dT}\right).
\end{equation}
Based on the relation in Eq.~\eqref{key relation}, the linear and solitonic-edge speeds, denoted as $s_-$ and $s_+$, of the cross sectional DSW along $m = n$ of the two-dimensional FPUT lattice \eqref{Main model} read:
\begin{equation}\label{cross-sectional DSW-fitting edge speeds}
    s_{\pm} = \frac{1}{2}\left(c_s + \epsilon^2S_{\pm}^{(X,T)}\right).
\end{equation}

\begin{figure}[b!]
    \centering
    \includegraphics[width=0.99\linewidth]{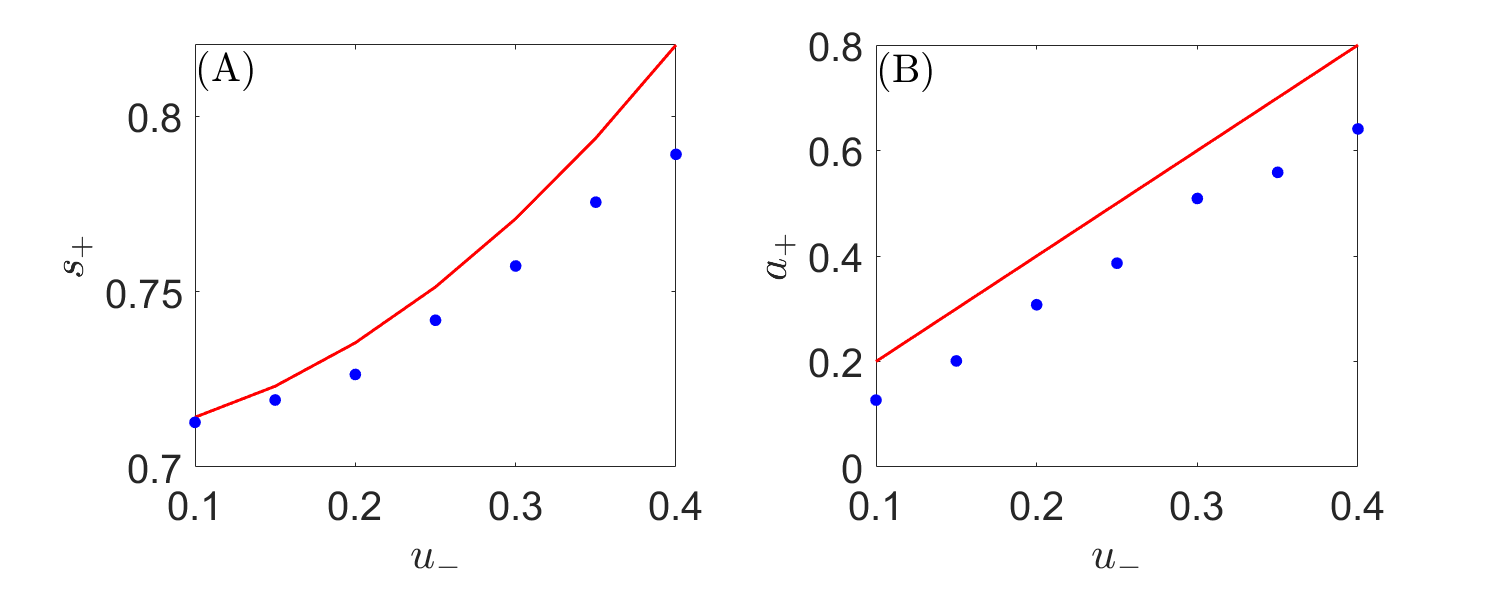}
    \caption{The comparison of the solitonic-edge features based on the DSW-fitting theoretical predictions and their associated numerical counterparts. Panels (A) and (B) depict the theoretically predicted (red curves) and numerically computed (blue dots) of the DSW solitonic-edge speeds and amplitudes along the direction of $m = n$.}
    \label{fig:soliton-edge features}
\end{figure}

In addition, we notice that the mKdV reduction in Eq.~\eqref{mKdV reduction} admits the following soliton solution, which is placed on the zero homogeneous background:
\begin{equation}\label{mKdV soliton solution}
    U(X,T) = \frac{\sqrt{2cc_s}}{\sqrt{K_3}}\sech\left(\frac{\sqrt{12c_sc}}{\sqrt{K_1}}\left(X-cT\right)\right),
\end{equation}
up to an unimportant phase shift, and $c$ denotes the speed of propagation of the soliton.
Then, based on the soliton solution in Eq.~\eqref{mKdV soliton solution}, the soliton amplitude-speed relation is given as follows:
\begin{equation}\label{mKdV soliton amp-s rela}
    a = \frac{\sqrt{2cc_s}}{\sqrt{K_3}},
\end{equation}
where $a$ is denoted as the soliton amplitude.

By replacing the speed of propagation of $c$ with the DSW-fitting theoretical prediction on the solitonic-edge speed of $S_+$, we have the following analytical prediction on the DSW soliton amplitude, denoted by $a_+$,
\begin{equation}\label{DSW-fitting prediction on amplitude}
    a_+ = \frac{\epsilon\sqrt{2c_sS_+^{(X,T)}}}{\sqrt{K_3}}.
\end{equation}
However, it is worth noticing that Eq.~\eqref{DSW-fitting prediction on amplitude} is valid only if $U_+ = 0$ as the soliton solution in Eq.~\eqref{mKdV soliton solution} is placed on the zero homogeneous background. In general, when $U_+ \neq 0$, it is necessary to compute the periodic traveling-wave solution associated with Eq.~\eqref{mKdV reduction} and derive the associated amplitude-speed relation to find the DSW-fitting theoretical prediction on the solitonic-edge amplitude of the DSW.

We shall compare these DSW theoretical predictions in Eqs.~\eqref{cross-sectional DSW-fitting edge speeds} and \eqref{DSW-fitting prediction on amplitude} with their corresponding numerical counterparts to examine the performance of DSW fitting. Before we display these relevant comparison results, we discuss in detail the numerical approaches to estimate the edge features of the DSW. Firstly, for the solitonic amplitude of the DSW, it can be simply computed by evaluating the global maximum of the numerical solution. Namely, if we denote the cross section along $m = n$ of the numerical solution of the lattice in Eq.~\eqref{Main model} as $\tilde{u}_n(t)$, then the numerical solitonic-edge amplitude is given as:
\begin{equation}\label{Numerical DSW soliton amplitude}
    a_+(t) = \max_{1\leq n \leq N}\{\tilde{u}_n(t)\},
\end{equation}
where $N\in\mathbb{Z}$ is the total number of lattice nodes utilized in the numerical simulation.

On the other hand, to compute the numerical edge speeds of the cross sectional DSW of Eq.~\eqref{Main model}, for each time snapshot of $t$, we evaluate the solitonic-edge location of $n_+$ with the following formula:
\begin{equation}\label{time-series for soliton location}
    n_+(t) = \argmax_{1\leq n \leq N}\{\tilde{u}_n(t)\}.
\end{equation}
Then, the numerical solitonic-edge speed is computed as follows:
\begin{equation}
    s_+ = \frac{n_+(t_2) - n_+(t_1)}{t_2 - t_1},
\end{equation}
where $t_1 < t_2$ are two distinct time snapshots.

Fig.~\ref{fig:soliton-edge features} delineates the solitonic-edge feature comparisons. Based on Panels of $(a)$ and $(b)$, we can clearly see that both the solitonic-edge speed and amplitude comparisons are reasonably well, despite the fact that discrepancy becomes more prominent as the value of the initial jump $\Delta \equiv u_- - u_+$ becomes greater. However, this discovery is not surprising, since it is expected that the regular DSW structure shall be gradually deformed so that the DSW fitting theoretical predictions become less reliable (See the ``entropy conditions" in Eq.~\eqref{Entropy conditions}).

\subsection{Self-similarity and rarefaction waves}

\begin{figure*}[t!]
    \centering
    \includegraphics[width=0.8\linewidth]{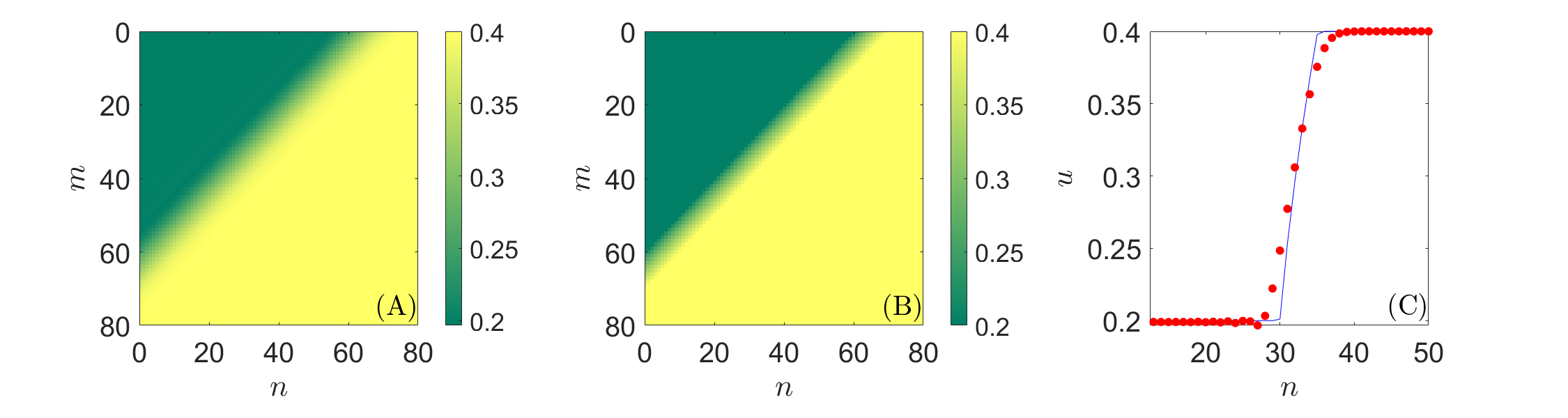}
    \caption{The comparison of the analytical self-similar solution in Eq.~\eqref{s.s solutions in n,m} with the numerical solution associated with the lattice in Eq.~\eqref{Main model} at $t = 40$. Panels (A), (B) depict the intensity plots associated with the numerical solution of Eq.~\eqref{Main model} and the self-similar solution in Eq.~\eqref{s.s solutions in n,m}, respectively. Panel (C) represents the cross-section plot of both solutions along the direction of $m = n$.}
    \label{Fig: rarefaction comparisons}
\end{figure*}

In this sub-section, we explore the analytical self-similar solutions associated with the dispersionless limit of the modified KdV reduction in \eqref{mKdV reduction}, which can be applied to approximate the rarefaction waves of the two-dimensional FPUT lattice in Eq.~\eqref{Main model}. It is worthwhile to note that when $K_{1,3} > 0$, the mKdV reduction Eq.~\eqref{mKdV reduction} is focusing and hence admits self-similar solutions \cite{el2017dispersive} which can model the spatial profiles of rarefactions. In particular, we first notice that the dispersionless limit of Eq.~\eqref{mKdV reduction} simply reads:
\begin{equation}\label{dispersionless limit of mKdV}
    U_T + \frac{3K_3}{c_s}U^2U_X = 0.
\end{equation}
Then, we look for self-similar solutions in the following form:
\begin{equation}\label{self-similar ansatz}
    U(X,T) = S(\kappa), \quad \kappa = \frac{X}{T}.
\end{equation}
Substituting of Eq.~\eqref{self-similar ansatz} into Eq.~\eqref{dispersionless limit of mKdV} and solving for the variable of $S$ yields the following solutions:
\begin{equation}\label{self-similar solutions}
    \begin{aligned}
        &U(X,T) = 
        \begin{cases}
            u_-, \quad X \leq \frac{3K_3u_-^2}{c_s}T,\\
            \sqrt{\frac{c_sX}{3K_3T}}, \quad \frac{3K_3u_-^2}{c_s}T < X < \frac{3K_3u_+^2}{c_s}T,\\
            u_+, \quad X \ge \frac{3K_3u_+^2}{c_s}T.
        \end{cases}
    \end{aligned}
\end{equation}
With the relations specified in Eq.~\eqref{mKdV cov} and transform the spatial-temporal coordinates of $(X,T)$ back into $(n,m,t)$ yields the following approximations to the numerical rarefaction waves of Eq.~\eqref{Main model}:
\begin{equation}\label{s.s solutions in n,m}
   \begin{aligned}
    &u_{n,m}(t) = 
    \begin{cases}
        \epsilon u_-, \quad n+m\leq s_-t,\\
        \epsilon\sqrt{\frac{c_s\left(n+m-c_st\right)}{3K_3\epsilon^2t}},\quad s_-t < n+m < s_+t,\\
        \epsilon u_+, \quad n+m \geq s_+t,
    \end{cases}
    \end{aligned}
\end{equation}
where $s_{\pm}(t) = \left(\frac{3K_3\epsilon^2u_{\pm}^2}{c_s} + c_s\right)t$.

Fig.~\ref{Fig: rarefaction comparisons} shows the two-dimensional spatial profiles (Panels (A) and (B)) and the comparison of the cross-sectional rarefaction waves (Panel (C)). We note that Panel (C) represents the cross-sectional rarefaction wave comparison along the direction of $m = n$, where the close alignment between the discrete rarefaction represented by discrete red dots and the analytical self-similar solution in Eq.~\eqref{self-similar solutions} of the mKdV reduction \eqref{mKdV reduction} demonstrates the validity of utilizing the self-similar solutions to approximate the spatial profiles of the numerical rarefactions.

\section{KPII wedge problems}\label{Sec: Wedge problems}

We will consider the Cauchy problem associated with KPII equation \eqref{KPII with unity} with the following wedge initial condition \cite{cdvf-xnfw}:
\begin{equation}\label{Wedge IC}
    U(\tilde{X},\tilde{Y},0) = H\left(f(\tilde{Y}) - \tilde{X}\right),
\end{equation}
where $H(\eta)$ denotes the Heaviside step function and $f(\tilde{Y}) = q_o\tilde{Y}$. 

\begin{figure}[t!]
    \centering
    \includegraphics[width=0.99\linewidth]{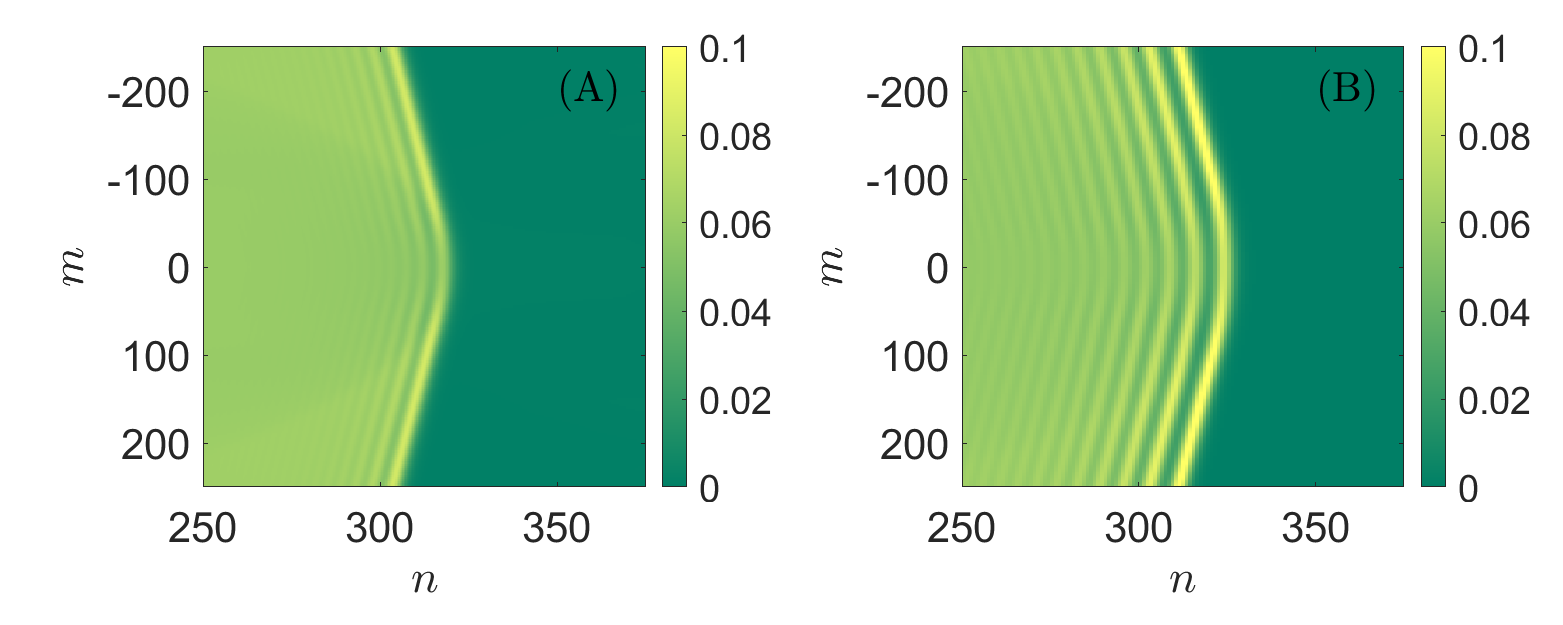}
    \caption{Comparison of the evolution dynamics at $t = 320$ of the subcritical type I wedge initial data. Panels (A) and (B) show the respective dynamics of the wedge initial data specified in Eq.~\eqref{Wedge IC} with $q_o = 0.4$ associated with the KPII reduction \eqref{KPII with unity} and the two-dimensional FPUT lattice \eqref{Main model}. Notice that in the numerical experiment, we set $\epsilon = 0.25$.}
    \label{fig:comparison for subcritical type-I wedge problem}
\end{figure}

\begin{figure}[b!]
    \centering
    \includegraphics[width=0.99\linewidth]{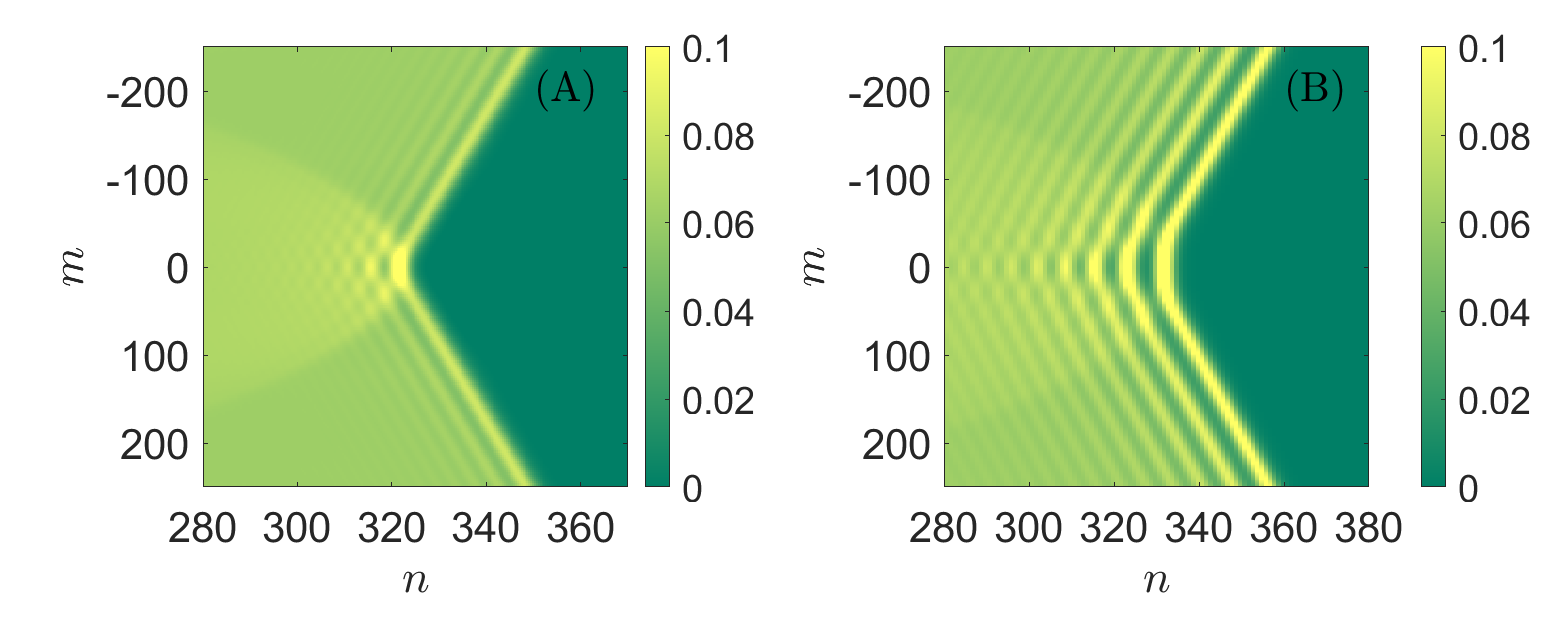}
    \caption{Same for Fig.~\ref{fig:comparison for subcritical type-I wedge problem} but now for the subcritical type II wedge initial data with $q_o = -0.7$.}
    \label{fig:comparison for subcritical type-II wedge problem}
\end{figure}

It is worthwhile to note that the value of the initial jump, denoted by $\Delta$, associated with the initial data in Eq.~\eqref{Wedge IC} reads $\Delta = 1$, and the critical angle that shall lead to distinct evolution dynamics of the wedge problem is $q_o^* = \sqrt{2}$. Specifically, this critical value specifies the parametric regime of $q$ for the wedge initial condition that will lead to qualitative different dynamics. In the present work, we consider only the two particular types of wedge initial data: subcritical I \& II \cite{cdvf-xnfw}. Fig.~\ref{fig:comparison for subcritical type-I wedge problem}-\ref{fig:comparison for subcritical type-II wedge problem} delineate the comparisons between the dynamics of the lattice \eqref{Main model} and the KPII reduction \eqref{KPII with unity} associated with the subcritical type I \& II initial wedge data, respectively. It can be clearly seen from these two comparisons for the dynamics associated with either the subcritical type I or type II wedge problems only admit qualitative agreements. In particular, there appears to exist a parametric delay in the KPII dynamics. For example, based on the dynamical comparison shown in Fig.~\ref{fig:comparison for subcritical type-I wedge problem}, the soliton edge in the expanding two-dimensional dispersive shock of the KPII reduction propagates with a slower speed that that of the lattice in Eq.~\eqref{Main model} does. 

\section{Conclusions and future challenges}\label{Sec: conclusions}

In the present work, we have derived two distinct long-wave asymptotic limits associated with an analog of the scalar two-dimensional FPUT lattice, and utilized the exact rational solutions of the mKdV equation to approximate the RW phenomena in the two-dimensional analog of the scalar FPUT lattice. Furthermore, we have also studied several dispersive patterns including the regular and compound dispersive structures and analytically captured their distinct edge characteristics via performing DSW fitting on the mKdV reduction, which has been equipped by the ``entropy conditions" to guarantee the validity of DSW fitting. Finally, we have also conducted numerical studies of the wedge problems associated with the lattice in Eq.~\eqref{Main model} and compared its evolution dynamics with that of the KPII reduction \eqref{KPII with unity}.

Although the present work conducts extensive numerical studies on different wave phenomena in the lattice \eqref{Main model}, there are lots of open questions. On the one hand, it is interesting to consider rogue-wave type solutions on top of periodic traveling-wave background specified in Ref.~\cite{mucalica2024dark,chen2018rogue}. The second direction is to apply DSW fitting directly on the lattice in Eq.~\eqref{Main model} based on the procedures specified in Ref.~\cite{chong2026travelingdispersiveshockwaves} to obtain edge-feature predictions on the discrete DSWs. Finally, we notice that the numerical comparisons conducted in Sec.~\ref{Sec: Wedge problems} admit only qualitative agreements delineated in Figs.~\ref{fig:comparison for subcritical type-I wedge problem} and \ref{fig:comparison for subcritical type-II wedge problem}. To obtain a better understanding on the numerical discrepancy on the dynamics of the wedge problems, it is important to investigate the two-dimensional DSW via Whitham analysis. Due to the integrability of the KPII reduction \eqref{KPII with unity}, it is possible to cast the Whitham modulation system \cite{ablowitz2017whithamKP} in its equivalent Riemann-invariant form and then take advantage of the diagonalized Whitham system to derive the spatially and temporally modulated periodic waves that can be regarded as the Whitham approximation to the numerical two-dimensional DSW. Moreover, the present paper has only considered the subcritical type I \& II wedge initial data, so one can also probe the other types of wedges mentioned in Ref.~\cite{cdvf-xnfw} (e.g. supercritical type I \& II). All these directions are currently under consideration and will be reported in future publications.

\bibliography{main}

\bibliographystyle{abbrv}

\end{document}